# Anomalous Hall transport by optically injected isospin degree of freedom in Dirac semimetal thin film

Yuta Murotani[1]*, Natsuki Kanda[1 †], Tomohiro Fujimoto[1], Takuya Matsuda[1 ‡], Manik Goyal[2], Jun Yoshinobu[1], Yohei Kobayashi[1], Takashi Oka[1], Susanne Stemmer[2], and Ryusuke Matsunaga[1]

[1]The Institute for Solid State Physics, The University of Tokyo, Kashiwa, Chiba 277-8581, Japan

[2]Materials Department, University of California, Santa Barbara, California 93106-5050, USA

ABSTRACT: Chirality of massless fermions emergent in condensed matter is a key to understand their characteristic behavior as well as to exploit their functionality. However, chiral nature of massless fermions in Dirac semimetals has remained elusive, due to equivalent occupation of carriers with the opposite chirality in thermal equilibrium. Here, we show that the isospin degree of freedom, which labels the chirality of massless carriers from a crystallographic point of view, can be injected by circularly polarized light. Terahertz Faraday rotation spectroscopy successfully detects the anomalous Hall conductivity by a light-induced isospin polarization in a three-dimensional Dirac semimetal, $Cd_3As_2$. Spectral analysis of the Hall conductivity reveals a long scattering time and a long decay time, which are characteristic of the isospin. The long-lived, robust, and reversible character of the isospin promises potential application of Dirac semimetals in future information technology.

Topological semimetals are of great interest as a platform of relativistic massless fermions[1,2], which give rise to several useful properties including ultrahigh mobility[3-5], broadband light detection[6-9], and large optical nonlinearity[10-15]. An essential characteristic of massless fermions is chirality, which distinguishes each particle from its mirror image. Because chirality is conserved in the ideal case, it is expected to serve as a robust information carrier in the next-generation nanoscale devices. In magnetic Weyl semimetals (WSMs), control of magnetic order enables electrical and optical switching of chirality[16,17], though the chirality itself does not play an important role there. Chiral anomaly provides another route to controlling chirality[18-21], but the necessity for a static magnetic field hinders fabrication of high-speed, integrated devices. To exploit the robustness of chirality, and to enhance its controllability, transport properties of chirality-polarized carriers should be addressed. Three-dimensional (3D) Dirac semimetals (DSMs) are a promising field of highly controllable chirality transport. In 3D DSMs, massless bands with the opposite chirality are completely degenerate in momentum space, so that no chirality polarization is present at thermal equilibrium. Such an initial state with no chirality polarization is suited for a plain background of information processing using chirality. In addition, the degenerate chirality degree of freedom may be useful in a manner similar to the degenerate spin degree of freedom in nonmagnetic materials, which has played an important role in charge-spin conversion in spintronics[22,23].

In real 3D DSMs such as $Na_3Bi$ and $Cd_3As_2$, the "isospin" degree of freedom takes the place of chirality. The two-fold isospin label distinguishes the overlapping massless bands based on their different irreducible representations in the crystallographic point group[18]. Figure 1a depicts the band structure of $Cd_3As_2$ as a combination of WSMs with the opposite isospins (up, ⇑ and down, ⇓)[24,25]. Two Dirac nodes appear on the $k_z$ axis because the four-fold rotational symmetry around the $c$ axis inhibits mixing of states with different irreducible representations. For the valley at $k_z > 0$ ($k_z < 0$), isospin up and down correspond to negative and positive (positive and negative) chirality of massless fermions, respectively. The opposite isospins are decoupled from each other in the low-energy region as shown in Supporting Information Note 1, which makes the isospin a more essential degree of freedom inherent to 3D DSMs. Because the isospin is correlated with the orbital angular momentum[24], one can expect that the optical selection rule enables injection of an isospin polarization by circularly polarized light (CPL). Figure 1b shows the isospin-resolved interband transition probability in momentum space, calculated by an effective four-band model of $Cd_3As_2$ for left-circularly polarized (LCP) light in the (112) plane. For right-circularly polarized (RCP) light, results for isospin up and down are interchanged. Details of calculation are presented in Supporting Information Note 1. At a low photon energy, imbalance between the opposite isospins is small because the dispersion is almost completely linear; carriers with momentum parallel or antiparallel to the propagation direction of light are mainly excited depending on the chirality[26,27] (left panel in Fig. 1b). As the photon energy increases, the two isoenergetic surfaces merge (middle panel in Fig. 1b) and approach a single sphere (right panel in Fig. 1b). In this regime, excitation in the isospin-up branch occurs mainly on the equator with respect to the propagation direction of light, while excitation in the isospin-down branch is limited to the poles. As a result, more electrons are excited in the isospin-up branch than in the isospin-down branch. This behavior is better seen in Fig. 1c, which shows the isospin-resolved absorption spectra and the resulting degree of isospin polarization. RCP light results in a degree of isospin polarization with the same magnitude but an opposite sign; linearly polarized light does not induce any isospin polarization. Thus, optical injection of an isospin polarization is expected, which may pave the way for unveiling the physics of isospin in 3D DSMs hidden at thermal equilibrium.

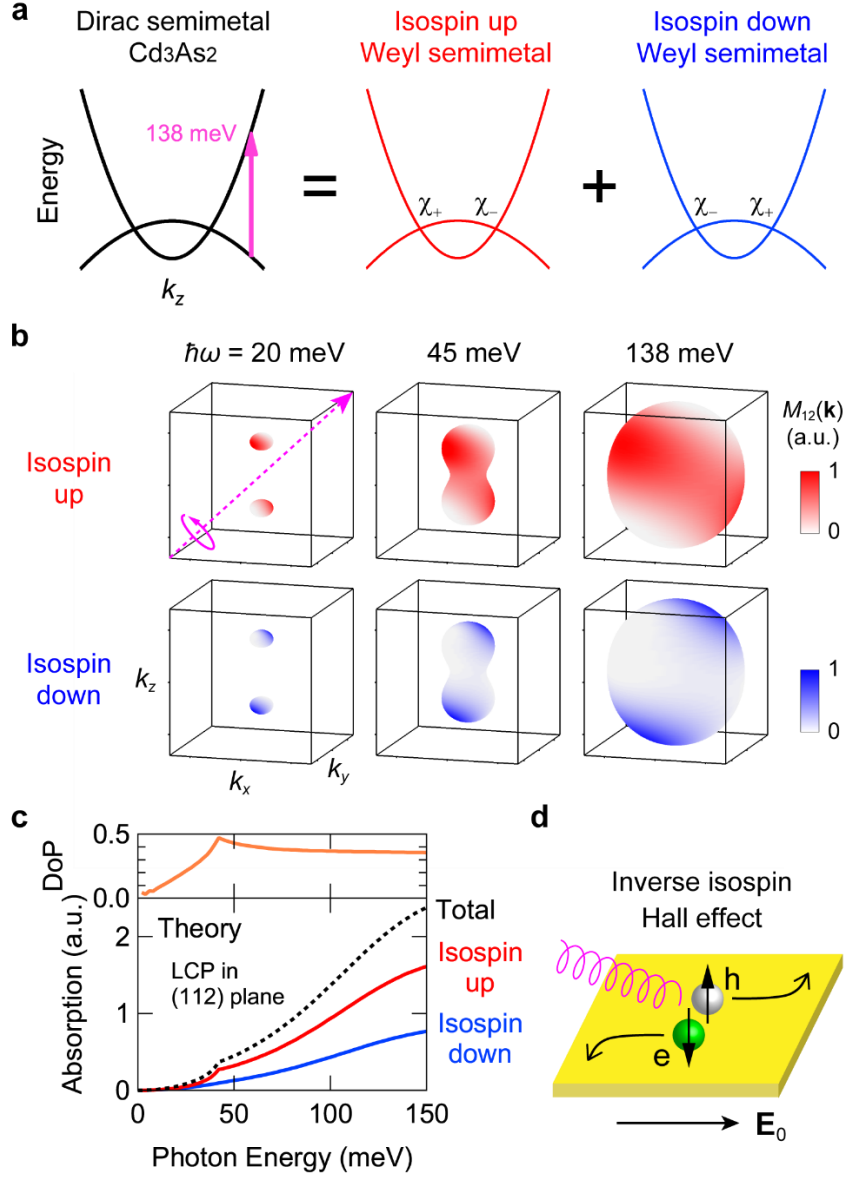

**Figure 1.** Optical injection of isospin in $Cd_3As_2$. (a) Dispersion relation of $Cd_3As_2$ along the $k_z$ axis. The doubly degenerate bands are distinguished by the isospin degree of freedom. $\chi_\pm$ labels the chirality of band-crossing points. (b) Interband transition probability $\mathcal{M}_{12}(\mathbf{k})$ for the LCP light polarized in the (112) plane of $Cd_3As_2$, with photon energies $\hbar\omega = 20$, 45, and 138 meV. The dashed arrow represents the propagation direction of light, i.e., $\hat{\mathbf{k}} \simeq (1/\sqrt{3})(1,1,1)$. a.u. stands for arbitrary units. (c) Bottom: Isospin-resolved interband absorption spectra for the LCP light polarized in the (112) plane of $Cd_3As_2$. Chemical potential $\mu = 88$ meV and temperature $T = 296$ K are assumed. Top: Degree of polarization (DoP) of the isospin, $(N_\Uparrow - N_\Downarrow)/(N_\Uparrow + N_\Downarrow)$, where $N_\Uparrow$ and $N_\Downarrow$ denote the excited carrier density of each isospin. A kink around 40 meV arises from a van-Hove singularity at the $\Gamma$ point. (d) Schematic of the inverse isospin Hall effect.

The light-induced isospin polarization is converted to a transverse electric current by an external electric field $\mathbf{E}_0$.

It is expected that an isospin polarization can be detected via the anomalous Hall effect, as is the case for spin, valley, and orbital polarizations in certain materials. Discovery of the inverse spin Hall effect, where a spin current is converted into a transverse electric current, greatly prompted the development of spintronics by allowing electrical detection of the spin polarization[22,23]. Valley polarizations in transition metal dichalcogenides can also be detected electrically through the valley Hall effect, paving the way for valleytronics[28,29]. Similar concepts of the orbital Hall effect and orbitronics have been proposed[30,31]. These kinds of Hall effect rely on time-reversal symmetry breaking by the spin, valley, and orbital polarizations. We can expect appearance of an inverse isospin Hall effect in 3D DSMs because the isospin polarization also breaks the time-reversal symmetry. Establishing the generation and detection methods for the isospin degree of freedom will greatly extend the functionality of 3D DSMs.

In this work, we demonstrate optical injection and detection of the isospin polarization in 3D DSMs. A $Cd_3As_2$ thin film at room temperature is excited by a circularly polarized multi-terahertz pump pulse. The resulting anomalous Hall conductivity is measured through polarization rotation of a single-cycle terahertz (THz) probe pulse. Spectral analysis of the anomalous Hall conductivity reveals a relatively long scattering time of carriers, which is attributed to the inverse isospin Hall effect caused by isospin-polarized holes in the neighborhood of Dirac nodes. Decay time of the light-induced anomalous Hall conductivity is also longer than the usual recombination lifetime, which is explained by a bottleneck effect around the Dirac nodes. These findings substantiate the long-lived nature of the isospin in 3D DSMs.

Figure 2a shows the experimental setup. An intense, circularly polarized, 150 fs-long multi-terahertz pump pulse[32,33] irradiates a high-quality, 240 nm-thick, (112)-oriented $Cd_3As_2$ thin film[34] at room temperature. Sample preparation and characterization are described in Supporting Information Note 2. The pump pulse with a photon energy of 138 meV selectively excites the low-energy Dirac bands (Fig. 1a) to inject an isospin polarization (Fig. 1c). To detect the inverse isospin Hall effect, we measure the polarization rotation of a linearly polarized THz probe pulse transmitting the sample[35,36] by electro-optic sampling with a compressed near-infrared pulse[37,38]. We achieve highly precise measurement of the polarization rotation angle reaching 0.1 mrad in the spectral range from 2 to 11 meV. For details of the experimental setup and spectral analysis[39], see Supporting Information Note 3. It should be noted that the temporal overlap between the pump and probe pulses leads to a complicated interplay of different mechanisms of transverse current generation, which has been investigated elsewhere[33]. Here we focus on the response after the pump pulse passes through the sample. Figure 2b shows the longitudinal conductivity spectrum $\sigma_{xx}(\omega)$ measured 2 ps after the pump irradiation with a fluence of 13 µJ/cm$^2$. We found that $\sigma_{xx}(\omega)$ is almost independent of the pump helicity, so here we present the average for the LCP and RCP pumps. The data are well fitted by the Drude model and is attributed to intraband absorption by pre-existing and photoexcited carriers[40,41]. Next, Fig. 2c shows the light-induced anomalous Hall conductivity $\Delta\sigma_{yx}(\omega)$ measured under the same condition. Here, half the difference between the results for LCP and RCP pumps is calculated, so as to extract the helicity-dependent signal. The spectrum is characterized by an opposite sign of the real and imaginary parts. Note, however, that when one assumes $\text{Im}\,\Delta\sigma_{yx}(\omega) > 0$ from the data, the Kramers-Kronig relation requires $\text{Re}\,\Delta\sigma_{yx}(0) > 0$, i.e., the real part must turn to a positive value in the DC limit. This expectation is supported by fitting with a theoretical curve, as explained below. Almost linear increase in the magnitude of $\Delta\sigma_{yx}$ against the excited carrier density $\Delta N$ is observed, as shown in Fig. 2d.

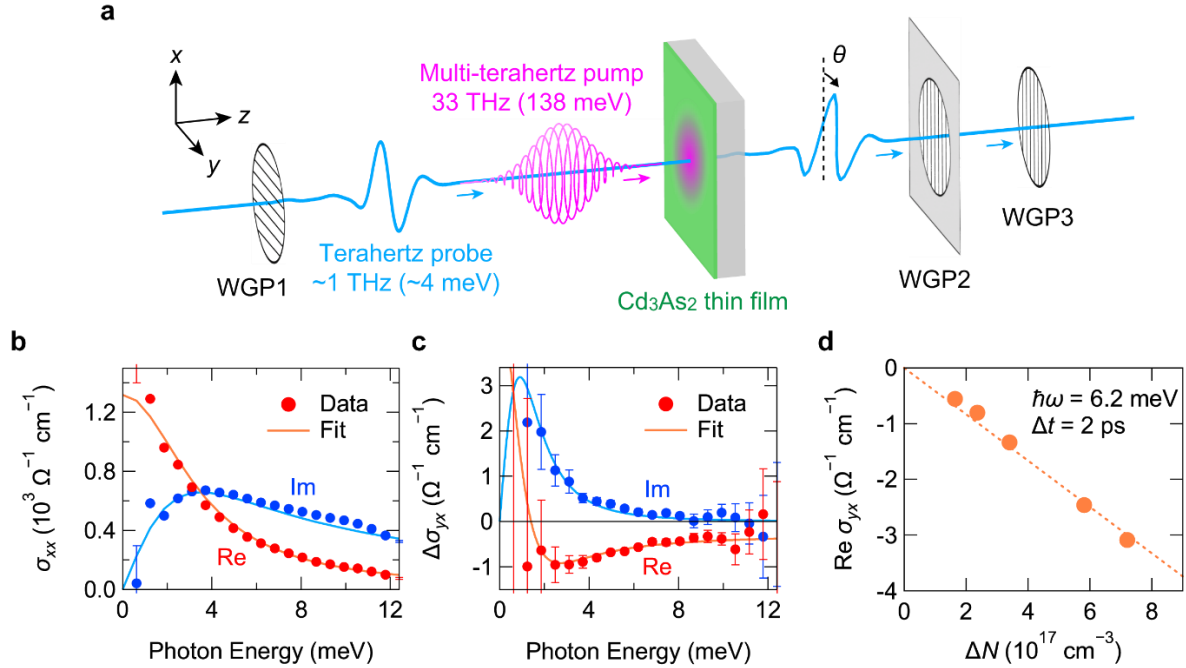

**Figure 2.** Anomalous Hall conductivity induced by photoexcitation. (a) Schematic of pump-probe experiment. WGP: wire grid polarizer. (b) Longitudinal conductivity spectrum $\sigma_{xx}(\omega)$ measured 2 ps after the pump irradiation with a fluence of 13 µJ/cm$^2$ (dots), along with a Drude model fitting (solid lines). The data are shown for (LCP pump + RCP pump)/2. Re and Im correspond to the real and imaginary parts, respectively. (c) Anomalous Hall conductivity spectrum $\Delta\sigma_{yx}(\omega)$ measured under the same conditions as (b), except for that the data are shown for (LCP pump − RCP pump)/2 (dots). The solid lines present the result of fitting by a model function including the skew scattering and intrinsic contributions. Error bars represent the statistical error. (d) Relation between the excited carrier density $\Delta N$ and the anomalous Hall conductivity at the probe photon energy $\hbar\omega = 6.2$ meV and $\Delta t = 2$ ps. $\Delta N$ is extracted from the Drude model fitting. The dashed line shows the linear relation. Error bars are smaller than dots.

To analyze the spectrum, we refer to a theoretical formula,

$$\Delta\sigma_{yx}(\omega) = \sum_{n=0}^{2} \frac{A_n}{(1-i\omega\tau)^n}, \tag{1}$$

where $n = 0, 1, 2$ correspond to the intrinsic, side jump, and skew scattering mechanisms of the anomalous Hall conductivity, respectively[42,43]. The intrinsic response arises from the Berry curvature in momentum space, which gives the photoexcited carriers the so-called anomalous velocity perpendicular to the external electric field[44]. Side jump is caused by a transverse shift of the carrier trajectory during impurity scattering[45]. Skew scattering also involves impurity scattering which changes the carrier momentum[46]. Spectral shape of each contribution is given in Supporting Information Note 4. Dashed lines in Fig. 2c show the result of fitting to the experimental data, taking into account the intrinsic ($n = 0$) and skew scattering ($n = 2$) contributions. Side jump ($n = 1$) is neglected because its inclusion does not improve the fitting. We find that the skew scattering well explains the overall behavior of the spectrum, including the opposite sign of the real and imaginary parts. The intrinsic contribution is found to be an order of magnitude smaller. In DC measurement, different mechanisms are often distinguished by different dependence on the scattering time: $A_0$ and $A_1$ should be independent of $\tau$, while $A_2$ should be proportional to $\tau$. In our experiment, it is not easy to vary $\tau$ by, e.g., degrading the sample or decreasing temperature. Frequency dependence of Eq. (1) serves as an alternative indicator of the microscopic mechanism, which is an advantage of our spectroscopic measurement[43]. Dominance of skew scattering is natural in view of the high quality of our sample, because it is known as the dominant origin of the AHE in clean ferromagnets[47,48].

From the fitting in Fig. 2c, we obtain $\tau_{\text{skew}} = 450$ fs as the scattering time for carriers contributing to the anomalous Hall conductivity. By contrast, the Drude model fitting in Fig. 2b yields a shorter scattering time, $\tau_{\text{Drude}} = 190$ fs. A large discrepancy between $\tau_{\text{Drude}}$ and $\tau_{\text{skew}}$ suggests that different kinds of carriers are responsible for the longitudinal ($\sigma_{xx}$) and transverse ($\sigma_{yx}$) transports. Effect of increasing pump fluence also differs for $\tau_{\text{Drude}}$ and $\tau_{\text{skew}}$. Figures 3a and 3b show the pump fluence dependence of $\sigma_{xx}$ and $\sigma_{yx}$ spectra, respectively, from which we extract the scattering times plotted in Fig. 3c. While $\tau_{\text{Drude}}$ stays almost constant, $\tau_{\text{skew}}$ appreciably decreases for increasing pump fluence, indicating that the anomalous Hall conductivity is more sensitive to the increase of carrier density.

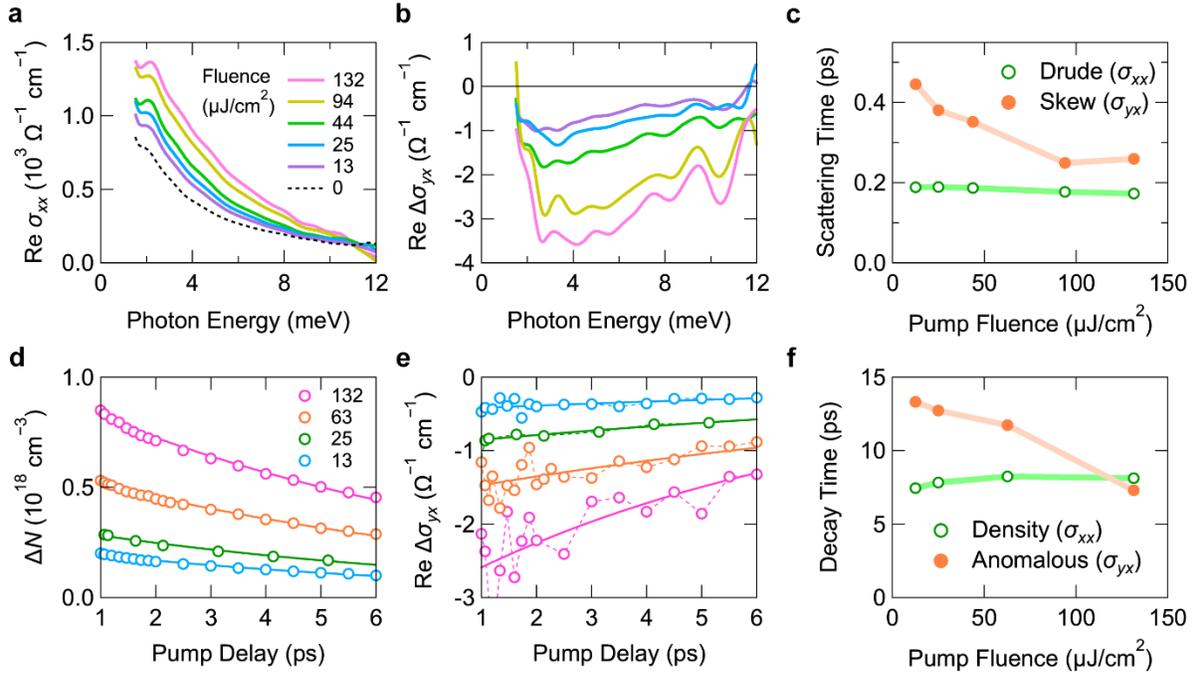

**Figure 3.** Pump fluence dependence. (a) Real part of the longitudinal optical conductivity $\sigma_{xx}(\omega)$ at $\Delta t = 2$ ps. (b) Real part of the optical Hall conductivity $\Delta\sigma_{yx}(\omega)$ at $\Delta t = 2$ ps. (c) Pump fluence dependence of $\tau_{\text{Drude}}$ (open circle) and $\tau_{\text{skew}}$ (solid circle). (d) Temporal change in the carrier density after pump irradiation, extracted from the Drude model fitting to $\sigma_{xx}(\omega)$. The thick lines show the results of exponential fitting. (e) Temporal change in Re $\Delta\sigma_{yx}(\omega)$ at $\hbar\omega = 6.2$ meV. The thick lines show the result of exponential fitting. (f) Pump fluence dependence of the carrier recombination time, $T_1$ (open circle), and the decay time of the anomalous Hall conductivity, $T_{\text{an}}$ (solid circle).

Figure 3d shows the pump delay dependence of $\Delta N$ extracted from $\sigma_{xx}$. Exponential fitting to this data reveals the lifetime of the excited carriers ($T_1$). On the other hand, Fig. 3e plots the pump delay dependence of Re $\Delta\sigma_{yx}$ at 6.2 meV, from which we can obtain the lifetime of the anomalous Hall conductivity ($T_{\text{an}}$). The results of fitting are combined in Fig. 3f. The recombination lifetime $T_1 = 7.7$ ps is consistent with previous studies[10,12,40,41] and remains almost constant for increasing pump fluence. By contrast, the anomalous Hall conductivity decays with a longer lifetime, which starts from $T_{\text{an}} = 13$ ps in the weak excitation limit and is shortened upon increasing the pump fluence. This contrast in the decay dynamics also supports our interpretation that $\sigma_{xx}$ and $\sigma_{yx}$ stem from different kinds of carriers.

In light of carrier relaxation dynamics depicted in Figs. 4a and 4b, we attribute the observed anomalous Hall conductivity to isospin-polarized holes in the vicinity of Dirac nodes. As shown by Fig. 1c, an LCP pump polarized in the (112) plane excites more electron-hole pairs in the isospin-up branch than in the isospin-down one. After photoexcitation, energy relaxation leads the photoexcited holes to be accumulated around the Dirac nodes, retaining a part of the isospin polarization (Fig. 4b). Isospin-polarized carriers are robust against scattering especially in the vicinity of Dirac nodes because a small density of states restricts the scattering paths[18], which accounts for the long scattering time $\tau_{\text{skew}} = 450$ fs in the anomalous Hall conductivity. Crystalline symmetry of DSMs also prevents the opposite isospins from mixing, which suppresses isospin relaxation[18]. A previous nonlocal transport experiment has estimated the isospin relaxation time $\tau_i$ as long as 200 ps in $Cd_3As_2$[19]. A different theoretical calculation has predicted $\tau_i = 50$ ps, for a Fermi level and a scattering time ($\tau_{\text{Drude}}$) close to ours[20]. For photoexcited carriers considered here, recombination of electron-hole pairs occurs simultaneously with simple isospin flipping, so that the decay time of the light-induced anomalous Hall conductivity ($T_{an}$) is shorter than the reported values of $\tau_i$. Decay rate of the anomalous Hall conductivity ($1/T_{an}$) should be given by a sum of the isospin relaxation rate ($1/\tau_i$) and the recombination rate for isospin-polarized carriers ($1/T_1^{isospin}$): $1/T_{an} = 1/\tau_i + 1/T_1^{isospin}$. If one assumes $T_1^{isospin} = T_1$, one obtains an unphysical, negative isospin relaxation rate for $T_{an} = 13$ ps and $T_1 = 7.7$ ps evaluated above. This inconsistency suggests that the recombination rate for the isospin-polarized carriers ($1/T_1^{isospin}$) is lower than the recombination rate for most carriers ($1/T_1$). Assuming $\tau_i = 50$ ps from the previous work[20], we obtain $T_1^{isospin} = 18$ ps, more than two times longer than $T_1$. We attribute the long lifetime of the isospin-polarized holes to a bottleneck effect around the Dirac nodes, where a small density of states suppresses scattering events. It is not clear how much degree of isospin polarization is retained in the course of initial relaxation to the Dirac nodes, because the magnitude of the anomalous Hall conductivity per an isospin-polarized hole is not known. To enable quantitative comparison with Fig. 1c, further theoretical consideration is necessary to take account of energy and momentum relaxation processes. Unlike holes, the excited electrons are relaxed mainly to the Fermi surface because of an unintentional n-type doping in our sample (Fig. 4b). They are expected to lose the information of isospin more rapidly because electronic states energetically away from the Dirac nodes undergo faster isospin relaxation[18]. At high excitation intensities, holes also occupy higher energy states where isospin relaxation is more efficient, which accounts for the decrease in $T_{an}$ in Fig. 3f. A larger density of states simultaneously increases the number of scattering paths for the anomalous Hall response, leading to a decrease in scattering time in Fig. 3c. On the other hand, carriers around the Fermi surface contributing to the longitudinal current are not much sensitive to the excitation intensity, which results in nearly constant scattering and decay times in Figs. 3c and 3f.

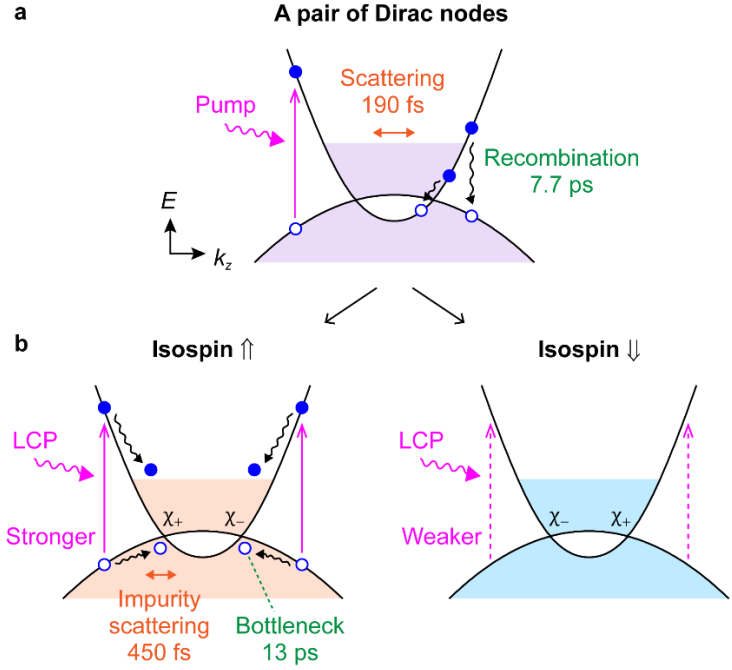

**Figure 4.** Relaxation dynamics of photoexcited carriers in $Cd_3As_2$. (a) After photoexcitation, most carriers feel a scattering time of $\tau_{\text{Drude}} = 190$ fs, and recombine with a lifetime of $T_1 = 7.7$ ps. (b) An LCP pump polarized in the (112) plane excites more electrons in the isospin up (⇑) branch than in the isospin down (⇓) branch, as shown by Fig. 1c. A part of the excited holes is accumulated in the vicinity of Weyl nodes in the course of energy relaxation, where they feel a scattering time of $\tau_{\text{skew}} = 450$ fs. The isospin-polarized holes vanish with a lifetime of $T_{\text{an}} = 13$ ps.

Since an isospin polarization is accompanied by a spin polarization[24], spin transport in DSMs may also benefit from the low scattering rate and protection by the crystalline symmetry. A robust spin transport in $Cd_3As_2$ has indeed been observed in a recent DC experiment with ferromagnetic contacts[49]. Unlike the conventional spin Hall effect in metals and semiconductors, the topological character of DSMs plays an important role in slowing down the relaxation of carriers, which makes the isospin more essential than the spin. Optical injection of a long-lived isospin polarization opens a novel route to chirality control in DSMs, that has been otherwise addressed only through chiral anomaly[19-21]. Here we would like to emphasize the advantage of measuring $\sigma_{yx}(\omega)$ in the frequency domain to study the isospin degree of freedom in DSMs. The longitudinal conductivity $\sigma_{xx}(\omega)$ mainly informs us of the majority carriers because it is contributed by all carriers in the system. On the other hand, the anomalous Hall conductivity spectrum $\sigma_{yx}(\omega)$ is sensitive to the carriers that break time-reversal symmetry, and its frequency dependence reflects the microscopic mechanism of the deflection of carrier trajectory. This feature makes it possible to reveal dynamics of isospin-polarized carriers.

To summarize, we experimentally demonstrated optical injection and detection of the isospin polarization in 3D DSMs. Using the time-resolved THz Faraday rotation spectroscopy, we experimentally determined the anomalous Hall conductivity spectrum of a $Cd_3As_2$ thin film excited by a circularly polarized multi-terahertz pump pulse. Spectral analysis revealed a relatively long scattering time $\tau_{skew} = 450$ fs for the carriers that participate in the transverse current. The relaxation time of the anomalous Hall conductivity, $T_{an} = 13$ ps, is found to be larger than the usual recombination lifetime $T_1 = 7.7$ ps, which is accounted for by a bottleneck effect in the vicinity of Dirac nodes. These findings extend the idea of spin Hall effect to the long-lived isospin degree of freedom in DSMs, and suggest a possibility of electrical readout of the optically induced isospin polarization, which may be integrated into future nanoscale devices.

ASSOCIATED CONTENT

**Supporting Information**.

Theory for isospin-resolved absorption spectrum, details of sample fabrication, experimental method, and fitting analysis of the data. (PDF)

AUTHOR INFORMATION

**Corresponding Author**

Yuta Murotani - *The Institute for Solid State Physics, The University of Tokyo, Kashiwa, Chiba 277-8581, Japan*; ORCID: https://orcid.org/0000-0003-3071-6408; E-mail: murotani@issp.u-tokyo.ac.jp

**ORCIDs:**

Natsuki Kanda: https://orcid.org/0000-0003-0512-6342

Tomohiro Fujimoto: https://orcid.org/0009-0008-7361-2725

Takuya Matsuda: https://orcid.org/0000-0003-0081-0743

Jun Yoshinobu: https://orcid.org/0000-0001-7774-8701

Yohei Kobayashi: https://orcid.org/0000-0002-9959-566X

Takashi Oka: https://orcid.org/0000-0003-1746-5368

Susanne Stemmer: https://orcid.org/0000-0002-3142-4696

Ryusuke Matsunaga: https://orcid.org/0000-0001-9674-5100

**Present Addresses**

†Research Center for Advanced Photonics, RIKEN, Wako, Saitama 351-0198, Japan.

‡Department of Physics, The University of Tokyo, Hongo, Tokyo 113-0033, Japan.

**Author Contributions**

YM and RM conceived this project. MG fabricated the sample with guidance from SS. NK and TM evaluated the linear response of the sample. YM, NK, and TF developed the pump-probe spectroscopy system with the help of JY, YK, and RM. YM performed the pump-probe experiment and analyzed the data with NK. YM, TO, and RM interpreted the results. YM conducted the theoretical calculations. All the authors discussed the data and interpretation. YM prepared the manuscript with substantial feedbacks from RM, SS, and all the other coauthors.

**Notes**

The authors declare no competing financial interest.

ACKNOWLEDGMENT

This work was supported by JST PRESTO (Grant Nos. JPMJPR20LA and JPMJPR2006), JST CREST (Grant No. JPMJCR20R4), and in part by JSPS KAKENHI (Grants Nos. JP20J01422 and JP20H00343). RM also acknowledges partial support by Attosecond lasers for next frontiers in science and technology (ATTO) in Quantum Leap Flagship Program (MEXT Q-LEAP). SS and MG acknowledge support by CATS Energy Frontier Research Center, which is funded by the Department of Energy, Basic Energy Sciences, under contract DE-AC02-07CH11358.

ABBREVIATIONS

CPL, circularly polarized light; DSM, Dirac semimetal; LCP, left-circularly polarized; RCP, right-circularly polarized; 3D, three-dimensional; THz, terahertz; WSM, Weyl semimetal; WGP, wire grid polarizer.

REFERENCES

1. Burkov, A. A. Topological semimetals. *Nat. Mater.* **2016**, 15, 1145-1148.

2. Armitage, N. P.; Mele, E. J.; Vishwanath, A. Weyl and Dirac semimetals in three-dimensional solids. *Rev. Mod. Phys.* **2018**, 90, 015001.

3. Liang, T.; Gibson, Q.; Ali, M. N.; Liu, M.; Cava, R. J.; Ong, N. P. Ultrahigh mobility and giant magnetoresistance in the Dirac semimetal $Cd_3As_2$. *Nat. Mater.* **2015**, 14, 280-284.

4. Shekhar, C.; Nayak, A. K.; Sun, Y.; Schmidt, M.; Nicklas, M.; Leermakers, I.; Zeitler, U.; Skourski, Y.; Wosnitza, J.; Liu, Z.; Chen, Y.; Schnelle, W.; Borrmann, H.; Grin, Y.; Felser, C.; Yan, B. Extremely large magnetoresistance and ultrahigh mobility in the topological Weyl semimetal candidate NbP. *Nat. Phys.* **2015**, 11, 645-649.

5. Huang, X.; Zhao, L.; Long, Y.; Wang, P.; Chen, D.; Yang, Z.; Liang, H.; Xue, M.; Weng, H.; Fang, Z.; Dai, X.; Chen, G. Observation of the Chiral-Anomaly-Induced Negative Magnetoresistance in 3D Weyl Semimetal TaAs. *Phys. Rev. X* **2015**, 5, 031023.

6. Wang, Q.; Li, C.-Z.; Ge, S.; Li, J.-G.; Lu, W.; Lai, J.; Liu, X.; Ma, J.; Yu, D.-P.; Liao, Z.-M.; Sun, D. Ultrafast Broadband Photodetectors Based on Three-Dimensional Dirac Semimetal $Cd_3As_2$. *Nano Lett.* **2017**, 17, 834-841.

7. Yang, M.; Wang, J.; Han, J.; Ling, J.; Ji, C.; Kong, X.; Liu, X.; Huang, Z.; Gou, J.; Liu, Z.; Xiu, F.; Jiang, Y. Enhanced Performance of Wideband Room Temperature Photodetector Based on $Cd_3As_2$ Thin Film/Pentacene Heterojunction. *ACS Photon.* **2018**, 5, 3438-3445.

*8.* Liu, J.; Xia, F.; Xiao, D.; García de Abajo, F. J.; Sun, D. Semimetals for high-performance photodetection. *Nat. Mater.* **2020**, 19, 830-837.

9. Zeng, L.; Han, W.; Ren, X.; Li, X.; Wu, D.; Liu, S.; Wang, H.; Lau, S. P.; Tsang, Y. H.; Shan, C.-X.; Jie, J. Uncooled Mid-Infrared Sensing Enabled by Chip-Integrated Low-Temperature-Grown 2D PdTe$_2$ Dirac Semimetal. *Nano Lett.* **2023**, 23, 8241-8248.

10. Zhu, C.; Wang, F.; Meng, Y.; Yuan, X.; Xiu, F.; Luo, H.; Wang, Y.; Li, J.; Lv, X.; He, L.; Xu, Y.; Liu, J.; Zhang, C.; Shi, Y.; Zhang, R.; Zhu, S. A robust and tuneable mid-infrared optical switch enabled by bulk Dirac fermions. *Nat. Commun.* **2017**, 8, 14111.

11. Meng, Y.; Zhu, C.; Li, Y.; Yuan, X.; Xiu, F.; Shi, Y.; Xu, Y.; Wang, F. Three-dimensional Dirac semimetal thin-film absorber for broadband pulse generation in the near-infrared. Opt. Lett. **2018**, 43, 1503−1506.

12. Cheng, B.; Kanda, N.; Ikeda, T. N.; Matsuda, T.; Xia, P.; Schumann, T.; Stemmer, S.; Itatani, J.; Armitage, N. P.; Matsunaga, R. Efficient Terahertz Harmonic Generation with Coherent Acceleration of Electrons in the Dirac Semimetal Cd$_3$As$_2$. *Phys. Rev. Lett.* **2020**, 124, 117402.

13. Kovalev, S.; Dantas, R. M. A.; Germanskiy, S.; Deinert, J.-C.; Green, B.; Ilyakov, I.; Awari, N.; Chen, M.; Bawatna, M.; Ling, J.; Xiu, F.; van Loosdrecht, P. H. M.; Surówka, P.; Oka. T.; Wang, Z. Non-perturbative terahertz high-harmonic generation in the three-dimensional Dirac semimetal Cd$_3$As$_2$. *Nat. Commun.* **2020**, 11, 2451.

14. Lim, J.; Ang, Y. S.; García de Abajo, F. J.; Kaminer, I.; Ang, L. K.; Wong, L. J. Efficient generation of extreme terahertz harmonics in three-dimensional Dirac semimetals. *Phys. Rev. Research* **2020**, 2, 043252.

15. Murotani, Y.; Kanda, N.; Ikeda, T. N.; Matsuda, T.; Goyal, M.; Yoshinobu, J.; Kobayashi, Y.; Stemmer, S.; Matsunaga R. Stimulated Rayleigh Scattering Enhanced by a Longitudinal

Plasma Mode in a Periodically Driven Dirac Semimetal $Cd_3As_2$. *Phys. Rev. Lett.* **2022**, 129, 207402.

16. Tsai, H.; Higo, T.; Kondou, K.; Kobayashi, A.; Nakano, T.; Yakushiji, K.; Miwa, S.; Otani, Y.; Nakatsuji, S. Spin–orbit torque switching of the antiferromagnetic state in polycrystalline $Mn_3Sn$/Cu/heavy metal heterostructures. AIP Adv. **2021**, 11, 045110.

17. Yoshikawa, N.; Ogawa, K.; Hirai, Y.; Fujiwara, K.; Ikeda, J.; Tsukazaki, A.; Shimano, R. Non-volatile chirality switching by all-optical magnetization reversal in ferromagnetic Weyl semimetal $Co_3Sn_2S_2$. Commun. Phys. **2022**, 5, 328.

18. Parameswaran, S. A.; Grover, T.; Abanin, D. A.; Pesin, D. A.; Vishwanath, A. Probing the Chiral Anomaly with Nonlocal Transport in Three-Dimensional Topological Semimetals. *Phys. Rev. X* **2014**, 4, 031035.

19. Zhang, C.; Zhang, E.; Wang, W.; Liu, Y.; Chen, Z.-G.; Lu, S.; Liang, S.; Cao, J.; Yuan, X.; Tang, L.; Li, Q.; Zhou, C.; Gu, T.; Wu, Y.; Zou, J.; Xiu, F.; Room-temperature chiral charge pumping in Dirac semimetals. *Nat. Commun.* **2017**, 8, 13741.

20. Cheng, B.; Schumann, T.; Stemmer, S.; Armitage, N. P. Probing charge pumping and relaxation of the chiral anomaly in a Dirac semimetal. *Sci. Adv.* **2021**, 7, eabg0914.

21. Park, B. C.; Ha, T.; Sim, K. I.; Jung, T. S.; Kim, J. H.; Kim, Y.; H. Lee, Y.; Kim, T.-T.; Kim, S. W. Real-space imaging and control of chiral anomaly induced current at room temperature in topological Dirac semimetal. *Sci. Adv.* **2022**, 8, eabq2479.

22. Žutić, I.; Fabian, J.; Das Sarma, S. Spintronics: Fundamentals and applications. *Rev. Mod. Phys.* **2022**, 76, 323-410.

23. Sinova, J.; Valenzuela, S. O.; Wunderlich, J.; Back, C. H.; Jungwirth, T. Spin Hall effects. *Rev. Mod. Phys.* **2015**, 87, 1213-1259.

24. Wang, Z.; Weng, H.; Wu, Q.; Dai, X.; Fang, Z. Three-dimensional Dirac semimetal and quantum transport in $Cd_3As_2$. *Phys. Rev. B* **2013**, 88, 125427.

25. Chen, R.; Wang, C. M.; Liu, T.; Lu, H.-Z.; Xie, X. C.; Quantum Hall effect originated from helical edge states in $Cd_3As_2$. *Phys. Rev. Research* **2021**, 3, 033227.

26. Yu, R.; Weng, H.; Fang, Z.; Ding, H.; Dai, X. Determining the chirality of Weyl fermions from circular dichroism spectra in time-dependent angle-resolved photoemission, *Phys. Rev. B* **2016**, 93, 205133.

27. Ma, Q.; Xu, S.-Y.; Chan, C.-K.; Zhang, C.-L.; Chang, G.; Lin, Y.; Xie, W.; Palacios, T.; Lin, H.; Jia, S.; Lee, P. A.; Jarillo-Herrero, P.; Gedik, N. Direct optical detection of Weyl fermion chirality in a topological semimetal, *Nat. Phys.* **2017**, 13, 842-847.

28. Xiao, D.; Liu, G.-B.; Feng, W.; Xu, X.; Yao, W. Coupled Spin and Valley Physics in Monolayers of $MoS_2$ and Other Group-VI Dichalcogenides. *Phys. Rev. Lett.* **2012**, 108, 196802.

29. Mak, K. F.; McGill, K. L.; Park, J.; McEuen, P. L. The valley Hall effect in $MoS_2$ transistors. *Science* **2014**, 344, 1489-1492.

30. Bernevig, B. A.; Hughes, T. L.; Zhang, S.-C. Orbitronics: The Intrinsic Orbital Current in p-Doped Silicon, *Phys. Rev. Lett.* **2005**, 95, 066601.

31. Go, D.; Jo, D.; Lee, H.-W.; Kläui, M.; Mokrousov, Yu. Orbitronics: Orbital currents in solids, *EPL* **2021**, 135, 37001.

32. Sell, A.; Leitenstorfer, A.; Huber, R. Phase-locked generation and field-resolved detection of widely tunable terahertz pulses with amplitudes exceeding 100 MV/cm. *Opt. Lett.* **2008**, 33, 2767-2769.

33. Murotani, Y.; Kanda, N.; Fujimoto, T.; Matsuda, T.; Goyal, M.; Yoshinobu, J.; Kobayashi, Y.; Oka, T.; Stemmer, S.; Matsunaga, R. Disentangling the competing mechanisms of light-induced anomalous Hall conductivity in three-dimensional Dirac semimetal. *Phys. Rev. Lett.* **2023**, 131, 096901.

34. Goyal, M.; Galletti, L.; Salmani-Rezaie, S.; Schumann, T.; Kealhofer, D. A.; Stemmer, S. Thickness dependence of the quantum Hall effect in films of the three-dimensional Dirac semimetal $Cd_3As_2$. *APL Mater.* **2018**, 6, 026105.

35. Matsuda, T.; Kanda, N.; Higo, T.; Armitage, N. P.; Nakatsuji, S.; Matsunaga, R. Room-temperature terahertz anomalous Hall effect in Weyl antiferromagnet $Mn_3Sn$ thin films. *Nat. Commun.* **2020**, 11, 909.

36. Kanda, N.; Konishi, K.; Kuwata-Gonokami, M. Terahertz wave polarization rotation with double layered metal grating of complimentary chiral patterns. *Opt. Express* **2007**, 15, 11117-11125.

37. Lu, C.-H.; Tsou, Y.-J.; Chen, H.-Y.; Chen, B.-H.; Cheng, Y.-C.; Yang, S.-D.; Chen, M.-C.; Hsu, C.-C.; Kung, A. H. Generation of intense supercontinuum in condensed media. *Optica* **2014**, 1, 400-406.

38. Kanda, N.; Ishii, N.; Itatani, J.; Matsunaga, R. Optical parametric amplification of phase-stable terahertz-to-mid-infrared pulses studied in the time domain. *Opt. Express* **2021**, 29, 3479-3489.

39. Kindt, J. T.; Schmuttenmaer, C. A. Theory for determination of the low-frequency time-dependent response function in liquids using time-resolved terahertz pulse spectroscopy. *J. Chem. Phys.* **1999**, 110, 8589-8596.

40. Lu, W.; Ling, J.; Xiu, F.; Sun, D. Terahertz probe of photoexcited carrier dynamics in the Dirac semimetal $Cd_3As_2$. *Phys. Rev. B* **2018**, 98, 104310.

41. Kanda, N.; Murotani, Y.; Matsuda, T.; Goyal, M.; Salmani-Rezaie, S.; Yoshinobu, J.; Stemmer, S.; Matsunaga, R. Tracking Ultrafast Change of Multiterahertz Broadband Response Functions in a Photoexcited Dirac Semimetal $Cd_3As_2$ Thin Film. *Nano Lett.* **2022**, 22, 2358-2364.

42. Nozières, P.; Lewiner, C. A simple theory of the anomalous Hall effect in semiconductors. *J. Physique* **1973**, 34, 901-915.

43. Fujimoto, T.; Kurihara, T.; Murotani, Y.; Tamaya, T.; Kanda, N.; Kim, C.; Yoshinobu, J.; Akiyama, H.; Kato, T.; Matsunaga, R. Observation of terahertz spin Hall conductivity spectrum in GaAs with optical spin injection. arXiv:2305.09155 [cond-mat.mtrl-sci]. May 16th, 2023. https://doi.org/10.48550/arXiv.2305.09155. (accessed 2023-12-18).

44. Karplus, R.; Luttinger, J. M.; Hall Effect in Ferromagnetics. *Phys. Rev.* **1954**, 95, 1154-1160.

45. Berger, L. Side-Jump Mechanism for the Hall Effect of Ferromagnets. *Phys. Rev. B* **1970**, 2, 4559-4566.

46. Smit, J. The spontaneous Hall effect in ferromagnetics II. *Physica* **1958**, 24, 39-51.

47. Onoda, S.; Sugimoto, N.; Nagaosa, N. Intrinsic Versus Extrinsic Anomalous Hall Effect in Ferromagnets. *Phys. Rev. Lett.* **2006**, 97, 126602.

48. Onoda, S.; Sugimoto, N.; Nagaosa, N. Quantum transport theory of anomalous electric, thermoelectric, and thermal Hall effects in ferromagnets. *Phys. Rev. B* **2008**, 77, 165103.

49. Stephen, G. M.; Hanbicki, A. T.; Schumann, T.; Robinson, J. T.; Goyal, M.; Stemmer, S.; Friedman, A. L. Room-Temperature Spin Transport in $Cd_3As_2$. *ACS Nano* **2021**, 15, 5459-5466.

# Supporting Information for

# Anomalous Hall transport by optically injected isospin degree of freedom in Dirac semimetal thin film

Yuta Murotani[1], Natsuki Kanda[1], Tomohiro Fujimoto[1],

Takuya Matsuda[1], Manik Goyal[2], Jun Yoshinobu[1], Yohei Kobayashi[1],

Takashi Oka[1], Susanne Stemmer[2], and Ryusuke Matsunaga[1]

[1]*The Institute for Solid State Physics, The University of Tokyo, Kashiwa, Chiba 277-8581, Japan*

[2]*Materials Department, University of California, Santa Barbara, California 93106-5050, USA*

### Note 1. Calculation of isospin-resolved absorption spectrum

The low-energy band structure of $Cd_3As_2$ can be described by an effective four-band model[S1],

$$H(\mathbf{k}) = \epsilon_0(\mathbf{k})\sigma_0\tau_0 + Ak_x\sigma_z\tau_x - Ak_y\sigma_0\tau_y + M(\mathbf{k})\sigma_0\tau_z$$

$$= \epsilon_0(\mathbf{k}) + \begin{pmatrix} M(\mathbf{k}) & Ak_+ & 0 & 0 \\ Ak_- & -M(\mathbf{k}) & 0 & 0 \\ 0 & 0 & M(\mathbf{k}) & -Ak_- \\ 0 & 0 & -Ak_+ & -M(\mathbf{k}) \end{pmatrix}, \quad (S1)$$

with $\epsilon_0(\mathbf{k}) = C_0 + C_1 k_z^2 + C_2(k_x^2 + k_y^2)$, $M(\mathbf{k}) = M_0 - M_1 k_z^2 - M_2(k_x^2 + k_y^2)$, and $k_\pm = k_x \pm ik_y$. In the first line, $\tau_i$ and $\sigma_i$ are Pauli matrices associated with the orbital and spin angular momentum, respectively. Importantly, the up- and down-spin states (the upper and lower $2 \times 2$ blocks in the second line) are completely decoupled; off-diagonal terms such as $\sigma_x\tau_x$ and $\sigma_y\tau_x$ are cubic with respect to $\mathbf{k}$, which allows us to neglect them in the low-energy region. This decoupling originates from the high crystalline symmetry of the material and suppresses spin relaxation even in the presence of impurity scattering. To indicate the relationship to the crystalline symmetry and the robustness against perturbation, the "isospin" degree of freedom has been introduced rather than simple spin. In the second line of Eq. (S1), the upper and lower $2 \times 2$ subspaces correspond to the isospin up and down, respectively. We use $C_0 = -0.0145$ eV, $C_1 = 10.59$ eV Å$^2$, $C_2 = 11.5$ eV Å$^2$, $M_0 = -0.0205$ eV, $M_1 = -18.77$ eV Å$^2$, $M_2 = -13.5$ eV Å$^2$, and $A = 0.889$ eV Å, taken from Ref. S2. Fermi's golden rule tells that the transition probability from an energy eigenstate $|u_m(\mathbf{k})\rangle$ to another one $|u_n(\mathbf{k})\rangle$ is proportional to

$$\mathcal{M}_{nm}(\mathbf{k}) = |\mathbf{e} \cdot \mathbf{J}_{nm}(\mathbf{k})|^2, \quad (S2)$$

where $\mathbf{e}$ is the polarization vector of the pump pulse, and $\mathbf{J}_{nm}(\mathbf{k}) = \langle u_n(\mathbf{k})|\mathbf{J}(\mathbf{k})|u_m(\mathbf{k})\rangle$ is the matrix element of the current operator $\mathbf{J}(\mathbf{k}) = (e/\hbar)\nabla_\mathbf{k} H(\mathbf{k})$. For the left-circularly polarized light normally incident on a (112)-oriented $Cd_3As_2$ film, the polarization vector is approximately given by $\mathbf{e} = (\mathbf{e}_X + i\mathbf{e}_Y)/\sqrt{2}$, with $\mathbf{e}_X = (1/\sqrt{2})(1,-1,0)$ and $\mathbf{e}_Y = (1/\sqrt{6})(1,1,-2)$. Figure 1b in the main text shows the calculated values of $\mathcal{M}_{nm}(\mathbf{k})$ on the surfaces with constant transition energies.

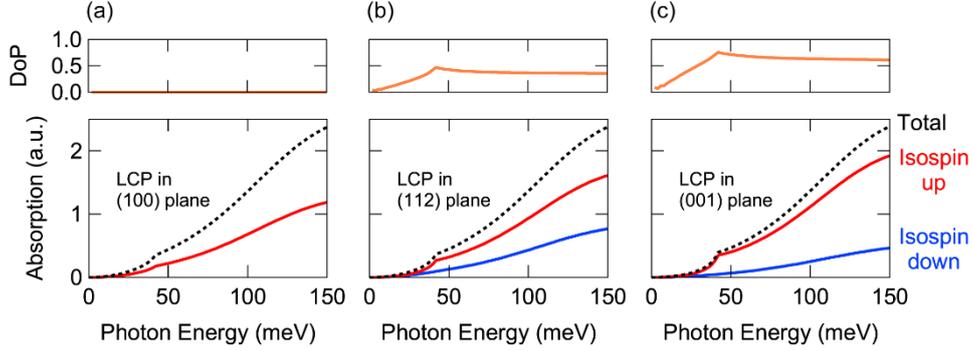

**Figure S1.** Theoretical calculation of isospin-dependent response. (a)-(c) Results for LCP light polarized in (100), (112), and (001) planes, respectively. Bottom: Absorption spectrum decomposed into isospin-up and isospin-down branches. Top: Degree of polarization (DoP) of the isospin, $(N_\Uparrow - N_\Downarrow)/(N_\Uparrow + N_\Downarrow)$, where $N_\Uparrow$ and $N_\Downarrow$ denote the excited carrier density of each isospin.

According to the Kubo formula, the absorption spectrum is given by

$$\sigma_{\text{abs}}(\omega) = \pi \int \frac{d^3k}{(2\pi)^3} \sum_{nm} \frac{f_m(\mathbf{k}) - f_n(\mathbf{k})}{\hbar\Omega_{nm}(\mathbf{k})} \mathcal{M}_{nm}(\mathbf{k}) \delta(\omega - \Omega_{nm}(\mathbf{k})), \qquad (S3)$$

where $f_n(\mathbf{k})$ is the Fermi distribution function for the $n$th band, and $\hbar\Omega_{nm}(\mathbf{k}) = \epsilon_n(\mathbf{k}) - \epsilon_m(\mathbf{k})$ is the transition energy. In Fig. S1, we calculated Eq. (S3) for the isospin-up and isospin-down branches assuming a chemical potential $\mu = 88$ meV above the Dirac nodes and a temperature $T = 296$ K. Difference between the isospin up and down branches vanishes for the (100) plane, leading to no degree of isospin polarization as shown in Fig. S1a. On the other hand, a large difference appears in the (001) plane, giving rise to an appreciable degree of isospin polarization as shown in Fig. S1c. The result for the (112) plane lies between those of the (100) and (001) planes, as seen in Fig. S1b. Figure 1c in the main text shows the same result as Fig. S1b.

### Note 2. Sample Preparation and characterization

A 240 nm-thick, (112)-oriented $Cd_3As_2$ thin film was grown by molecular beam epitaxy on a GaAs substrate with a GaSb buffer layer[S3]. A relatively low Fermi level was confirmed by the onset of interband transitions as low as 117 meV, demonstrating the high quality of the sample[S4].

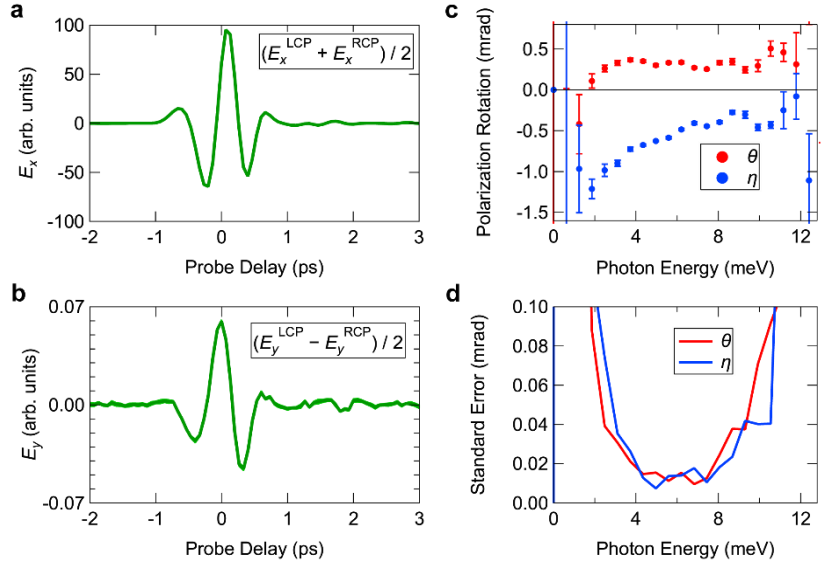

**Figure S2.** Example of measured waveforms and polarization rotation. (a) $E_x$ waveform for pump fluence 13 µJ/cm² and pump delay 2 ps. The average for the LCP and RCP pumps is shown. Statistical error is smaller than the line thickness. (b) $E_y$ waveform. Half the difference between the LCP and RCP pumps is shown. Statistical error is given by shaded curves. (c) Real and imaginary parts of the complex rotation angle $\theta(\omega) + i\eta(\omega) = E_y(\omega)/E_x(\omega)$. (d) Statistical error in the polarization rotation. Error bars in (c) show the same quantity.

A Drude model fit to the optical conductivity in the THz frequency region also shows an electron momentum relaxation time as long as $\tau_{\text{Drude}} = 190$ fs at room temperature, indicating a low concentration of impurities. Since the band gaps of GaAs (1.42 eV) and GaSb (0.73 eV) are much smaller than the pump photon energy (0.14 meV), the layers other than $Cd_3As_2$ were not excited by the pump pulse, which was confirmed by a measurement for a sample composed of a GaAs substrate and a GaSb layer only.

### Note 3. Multi-terahertz pump and terahertz Faraday rotation spectroscopy

We used a Yb:KGW-based regenerative amplifier as a light source (wavelength 1030 nm, pulse width 255 fs, pulse energy 2 mJ, and repetition rate 3 kHz). A large portion of the output beam pumped two optical parametric amplifiers providing signal beams at 1500 nm and 1800 nm. They were mixed in a 500 µm-thick GaSe crystal to generate a multi-terahertz pump pulse with frequency of 33.3 THz, pulse width of 150 fs, and pulse energy of 2 µJ, through difference frequency generation[S5,S6]. A commercially available quarter waveplate was used to make it circularly polarized. We defined the left circular polarization as a counterclockwise rotation of the electric field viewed from the propagation direction, and the right circular polarization inversely.

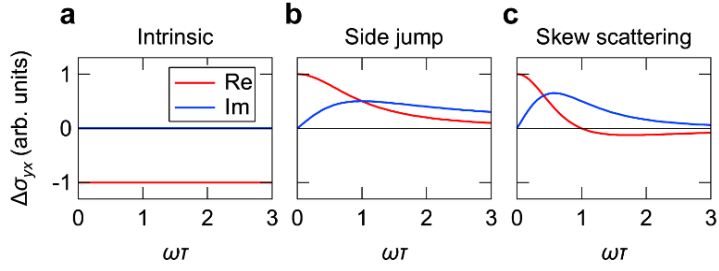

**Figure S3.** Theoretical model for anomalous Hall conductivity. (a), (b), (c) Contributions from the intrinsic, side jump, and skew scattering mechanisms, respectively. The formula is given by Eq. (1) in the main text.

The remaining part of the light source was compressed down to 80 fs with a multi-plate broadening scheme[S7,S8] and was divided into two beams. The stronger one generated the THz probe pulse through optical rectification in a 2 mm-thick (110) GaP crystal. The weaker one was used as the gate pulse which detected the THz pulse by electro-optic sampling. A combination of rotatable and fixed wire grid polarizers after the sample enabled determination of the polarization state of the transmitted probe[S9,S10]. Both the pump and probe pulses were normally incident on the sample with spot sizes ($1/e^2$ diameter) of 1.6 mm and 0.7 mm, respectively. Their arrival times are denoted by $t_{\text{pump}}$ and $t_{\text{probe}}$, while the gate pulse measured the probe electric field at the time $t_{\text{gate}}$. We defined $\Delta t = t_{\text{gate}} - t_{\text{pump}}$ as the pump delay time, and performed Fourier transform with respect to the probe delay time $t_{\text{gate}} - t_{\text{probe}}$ to obtain frequency dependence of response functions[S11].

Figures S2(a) and (b) show the $E_x$ and $E_y$ waveforms obtained for a pump fluence of 13 µJ/cm$^2$ and pump delay 2 ps. Owing to the high stability of the laser, statistical error is invisible in the plots. Figure S2(c) shows the real and imaginary parts of the complex rotation angle $\theta(\omega) + i\eta(\omega) = E_y(\omega)/E_x(\omega)$ obtained from the waveforms in Figs. S2(a) and (b). Standard error in repeated measurement is shown as error bars, and also is given in Fig. S2(d). Precision of the polarization rotation angle reaches 0.1 mrad for 2-11 meV (0.02 mrad for 4-8 meV) within 2.5 h. The procedure to calculate the anomalous Hall conductivity spectrum from the complex polarization rotation angle is given in Ref. S12.

At present, it is difficult to examine pump photon energy dependence in Figs. 1b and 1c because of limitations in multi-terahertz optics; as for the light source, it is difficult to generate intense multi-terahertz pulses below 10 THz (41 meV), where the degree of isospin polarization is expected to decrease appreciably. Quarter waveplates are also scarce in this frequency region. Even if we could generate such circularly polarized pump pulses, intraband absorption neglected in Fig. 1c should complicate the situation. Development of optical elements in the multi-terahertz frequency region and electrical tuning of the Fermi energy may enable experimental access to the photon energy dependence characteristic to optically injected isospin.

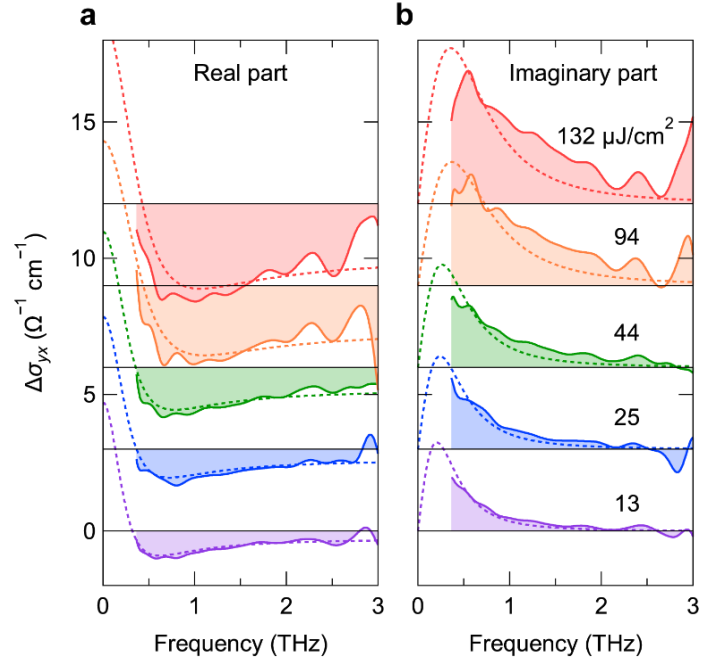

**Figure S4.** Fitting to the anomalous Hall conductivity at $\Delta t = 2$ ps. (a), (b) Real and imaginary parts respectively (solid lines). Dashed lines show the fitting results.

**Note 4. Fitting analysis**
Figure S3 compares the theoretical anomalous Hall conductivity for the intrinsic, side jump, and skew scattering mechanisms, given by Eq. (1) in the main text. Figure S4 shows the measured spectra for different pump intensities, along with the fitting result. We find that inclusion of a side jump contribution does not improve the fitting, so that only the intrinsic and skew scattering ones are taken into account here. The experimental data and the fitting results agree well, especially at low intensities. The spectrum at high intensities may be affected by carrier-carrier scattering.

Figure S5 shows the exponential fitting to the carrier density (panel (a)) and the anomalous Hall conductivity (panel (b)) dependent on the pump delay, for different pump intensities. The pump pulse reflects off the back surface of the sample, which causes repeated excitations at $\Delta t = 0, 7, 14, \ldots$ ps. Such multiple reflection prevents a fitting over a longer temporal range. We have confirmed that the decay dynamics after the second excitation is consistent with the decay times after the first excitation.

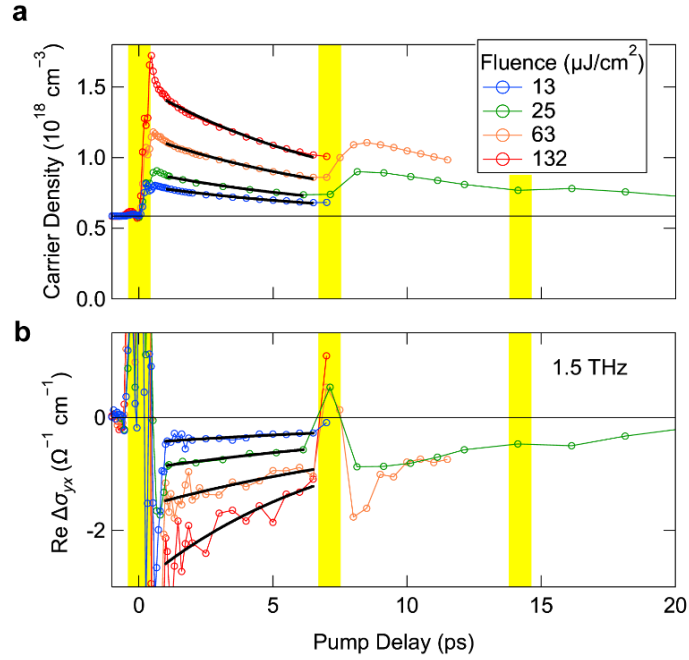

**Figure S5.** Decay dynamics. (a) Temporal change in the carrier density after pump irradiation, extracted from the Drude model fitting to $\sigma_{xx}$. Shaded regions show the arrival of the pump pulse, which occurs repeatedly due to multiple reflection inside the sample. The thick lines show the results of exponential fitting. (b) Temporal change in $\mathrm{Re}\,\Delta\sigma_{yx}$ at 1.5 THz. The thick lines show the results of exponential fitting.

**References**

[S1] Z. Wang, H. Weng, Q. Wu, X. Dai, and Z. Fang, Three-dimensional Dirac semimetal and quantum transport in $Cd_3As_2$, *Phys. Rev. B* **88**, 125427 (2013).

[S2] R. Chen, C. M. Wang, T. Liu, H.-Z. Lu, and X. C. Xie, Quantum Hall effect originated from helical edge states in $Cd_3As_2$, *Phys. Rev. Research* **3**, 033227 (2021).

[S3] M. Goyal, L. Galletti, S. Salmani-Rezaie, T. Schumann, D. A. Kealhofer, and S. Stemmer, Thickness dependence of the quantum Hall effect in films of the three-dimensional Dirac semimetal $Cd_3As_2$, *APL Mater.* **6**, 026105 (2018).

[S4] N. Kanda, Y. Murotani, T. Matsuda, M. Goyal, S. Salmani-Rezaie, J. Yoshinobu, S. Stemmer, and R. Matsunaga, Tracking Ultrafast Change of Multiterahertz Broadband Response Functions in a Photoexcited Dirac Semimetal $Cd_3As_2$ Thin Film, *Nano Lett.* **22**, 2358 (2022).

[S5] A. Sell, A. Leitenstorfer, and R. Huber, Phase-locked generation and field-resolved detection of widely tunable terahertz pulses with amplitudes exceeding 100 MV/cm, *Opt. Lett.* **33**, 2767 (2008).

[S6] Y. Murotani, N. Kanda, T. N. Ikeda, T. Matsuda, M. Goyal, J. Yoshinobu, Y. Kobayashi, S. Stemmer, and R. Matsunaga, Stimulated Rayleigh Scattering Enhanced by a Longitudinal Plasma Mode in a Periodically Driven Dirac Semimetal $Cd_3As_2$, *Phys. Rev. Lett.* **129**, 207402 (2022).

[S7] C.-H. Lu, Y.-J. Tsou, H.-Y. Chen, B.-H. Chen, Y.-C. Cheng, S.-D. Yang, M.-C. Chen, C.-C. Hsu, and A. H. Kung, Generation of intense supercontinuum in condensed media, *Optica* **1**, 400 (2014).

[S8] N. Kanda, N. Ishii, J. Itatani, and R. Matsunaga, Optical parametric amplification of phase-stable terahertz-to-mid-infrared pulses studied in the time domain, *Opt. Express* **29**, 3479 (2021).

[S9] T. Matsuda, N. Kanda, T. Higo, N. P. Armitage, S. Nakatsuji, and R. Matsunada, Room-temperature terahertz anomalous Hall effect in Weyl antiferromagnet $Mn_3Sn$ thin films, *Nat. Commun.* **11**, 909 (2020).

[S10] N. Kanda, K. Konishi, and M. Kuwata-Gonokami, Terahertz wave polarization rotation with double layered metal grating of complimentary chiral patterns, *Opt. Express* **15**, 11117 (2007).

[S11] J. T. Kindt and C. A. Schmuttenmaer, Theory for determination of the low-frequency time-dependent response function in liquids using time-resolved terahertz pulse spectroscopy, *J. Chem. Phys.* **110**, 8589-8596 (1999).

[S12] Y. Murotani, N. Kanda, T. Fujimoto, T. Matsuda, M. Goyal, J. Yoshinobu, Y. Kobayashi, T. Oka, S. Stemmer, and R. Matsunaga, Disentangling the competing mechanisms of light-induced anomalous Hall conductivity in three-dimensional Dirac semimetal, *Phys. Rev. Lett.* **131**, 096901 (2023).